\documentstyle[12pt,axodraw]{article}

\newcommand{\re}{\mbox{$\rm e$}}
\newcommand{\ri}{\mbox{$\rm i$}}

\newcommand{\sd}{\mbox{$\Sigma^\dagger$}}

\newcommand{\xid}{\mbox{$\xi^\dagger$}}

\newcommand{\bfm}[1]{\mbox{\boldmath$#1$}}
\newcommand{\ratio}[2]{\mbox{$#1\over#2$}}
\newcommand{\tr}{\mbox{$\,\rm Tr$}}

\newcommand{\ra}{\rightarrow}

\begin{document}
\baselineskip=17pt
\parskip=5pt

\begin{titlepage}
\vfill

\hskip 4in {ISU-HET-98-4}

\hskip 4in {September 1998}
\vspace{1 in}
\begin{center}
{\large\bf $|\Delta\bfm{I}|=3/2$ Decays of the  $\Omega^-$ in Chiral
Perturbation Theory}\\ 

\vspace{1 in}
{\bf Jusak~Tandean}  
{\bf  and G.~Valencia}\\
{\it           Department of Physics and Astronomy,
               Iowa State University,
               Ames IA 50011}\\
\vspace{1 in}
\end{center}
\begin{abstract}
 
We study the decays  $\,\Omega^-\ra\Xi\pi\,$  using heavy-baryon chiral 
perturbation theory to quantify the $\,|\Delta\bfm{I}|=1/2\,$ 
rule in these decay modes.  
The ratio of $\,|\Delta\bfm{I}|=3/2\,$ to $\,|\Delta\bfm{I}|=1/2\,$ amplitudes 
is somewhat larger in  these  decays than it is in other hyperon decays.  
At leading order there are two operators responsible for the 
$\,|\Delta\bfm{I}|=3/2\,$  parts of the  $\Omega^-$  decays which also 
contribute at one loop to other hyperon decays.  
These one-loop contributions are sufficiently
large to indicate (albeit not definitely) that the measured ratio 
$\,\Gamma(\Omega^-\ra\Xi^0\pi^-)/\Gamma(\Omega^-\ra\Xi^-\pi^0)\approx 2.7\,$ 
may be too large.

\end{abstract}

\end{titlepage}

\clearpage

\section{Introduction}

For a purely  $\,|\Delta\bfm{I}|=1/2\,$  weak interaction,  the ratio of 
decay rates  
$\,\Gamma(\Omega^-\ra\Xi^0\pi^-)/\Gamma(\Omega^-\ra\Xi^-\pi^0)\,$ 
would be 2.   
Instead, this ratio is measured to be approximately 2.7~\cite{pdb},  
and it has been claimed in the literature that this could signal  
a violation of the  $\,|\Delta\bfm{I}|=1/2\,$  rule~\cite{geocar}.

In this paper we construct the lowest-order chiral
Lagrangian that contributes to the  $\,|\Delta\bfm{I}|=3/2\,$ 
non-leptonic decays of the  $\Omega^-$  and extract information on the 
couplings by fitting the observed decay rates.\footnote{%
For the  $\,|\Delta\bfm{I}|=1/2\,$  sector, theoretical calculation 
to one loop has recently been done in Ref.~\cite{ems}.} 
We then compute the one-loop contributions of this Lagrangian to the
$\,|\Delta\bfm{I}|=3/2\,$  amplitudes in (octet) hyperon non-leptonic decays. 
We find that the measured  $\,|\Delta\bfm{I}|=3/2\,$  amplitude 
in  $\Omega^-$  decays is sufficiently large to be in conflict with the 
measured  $\,|\Delta\bfm{I}|=3/2\,$  amplitudes in (octet) hyperon
non-leptonic decays.  
This is not a definite conclusion because the combinations of couplings 
that appear in the two cases are different.

\section{Chiral Lagrangian}

The chiral Lagrangian that describes the interactions of the lowest-lying
mesons and baryons has been discussed extensively in the 
literature~\cite{baryon,manjen,hungary}. 
It is written down in terms of the  $3\times 3$  matrices  $\phi$  and  $B$  
which represent the pseudoscalar-meson and baryon octets, 
and of the Rarita-Schwinger tensor  $T_{abc}^\mu$   which describes 
the spin-$3/2$ baryon decuplet (we use the notation of Ref.~\cite{atv}).  
The octet pseudo-Goldstone bosons enter through the exponential  
$\,\Sigma=\exp({\ri}\phi/f).\,$  
The field $T_{abc}^\mu$  satisfies the constraint 
$\,\gamma_\mu^{}T^\mu_{abc}=0 \,$  and is completely symmetric in its SU(3) 
indices,  $a,b,c$~\cite{hungary}.   
Its components are (with the Lorentz index suppressed)     
\begin{eqnarray}  
\begin{array}{c} \displaystyle      
T_{111}^{}  \;=\;  \Delta^{++}   \;, \hspace{2em}   
T_{112}^{}  \;=\;  \ratio{1}{\sqrt{3}}\, \Delta^+   \;, \hspace{2em}    
T_{122}^{}  \;=\;  \ratio{1}{\sqrt{3}}\, \Delta^0   \;, \hspace{2em}    
T_{222}^{}  \;=\;  \Delta^-   \;,   
\vspace{2ex} \\   \displaystyle       
T_{113}^{}  \;=\;  \ratio{1}{\sqrt{3}}\, \Sigma^{*+}   \;, \hspace{2em}    
T_{123}^{}  \;=\;  \ratio{1}{\sqrt{6}}\, \Sigma^{*0}   \;, \hspace{2em}    
T_{223}^{}  \;=\;  \ratio{1}{\sqrt{3}}\, \Sigma^{*-}   \;, 
\vspace{2ex} \\   \displaystyle       
T_{133}^{}  \;=\;  \ratio{1}{\sqrt{3}}\, \Xi^{*0}   \;, \hspace{2em}    
T_{233}^{}  \;=\;  \ratio{1}{\sqrt{3}}\, \Xi^{*-}   \;, \hspace{2em}    
T_{333}^{}  \;=\;  \Omega^-   \;. 
\end{array}    
\end{eqnarray}      
Under chiral SU(3$)_{\rm L}^{}\times~$SU(3$)_{\rm R}^{}$,   
these fields transform as   
\begin{eqnarray}    
\Sigma  \;\rightarrow\;  L \Sigma R^\dagger   \;, \hspace{2em}   
B  \;\rightarrow\;  U B U^\dagger             \;, \hspace{2em}   
T_{abc}^\mu  \;\rightarrow\;  
U_{ad}^{} U_{be}^{} U_{cf}^{} T_{def}^\mu    \;,    
\end{eqnarray}      
where  $\,L,R\in\,$SU$(3)_{\rm L,R}^{}$  and the matrix  $U$  is 
implicitly defined by the transformation  
\begin{eqnarray}    
\xi  \;\equiv\;  \re^{{\rm i}\phi/(2f)}  
\;\rightarrow\;  L\xi U^\dagger \;=\;  U\xi R^\dagger   \;.    
\end{eqnarray}      

In the heavy-baryon formalism~\cite{manjen}, the effective Lagrangian is 
rewritten in terms of velocity-dependent baryon fields, 
$B_v^{}$  and  $T_v^\mu$.   
The leading-order chiral Lagrangian that describes the strong interactions 
of the pseudoscalar-meson and baryon octets as well as the baryon  
decuplet is given  by~\cite{manjen,hungary}
\begin{eqnarray}   \label{L1strong}
{\cal L}^{\rm s}  &\!\!\!=&\!\!\!      
\ratio{1}{4} f^2\,
 \tr \!\left( \partial^\mu \Sigma^\dagger\, \partial_\mu \Sigma \right) 
\;+\;  
\tr \!\left( \bar{B}_v^{}\, \ri v\cdot {\cal D} B_v^{} \right)    
\nonumber \\ && \!\!\! \!\!\!   
+\; 
2 D\, \tr \!\left( \bar{B}_v^{}\, S_v^\mu 
 \left\{ {\cal A}_\mu^{}\,,\, B_v^{} \right\} \right)     
+ 2 F\, \tr \!\left( \bar{B}_v^{}\, S_v^\mu\,    
 \left[ {\cal A}_\mu^{}\,,\, B_v^{} \right] \right)   
\nonumber \\ && \!\!\! \!\!\!   
-\;    
\bar{T}_v^\mu\, \ri v\cdot {\cal D} T_{v\mu}^{}  
+ \Delta m\, \bar{T}_v^\mu T_{v\mu}^{}  
+ {\cal C} \left( \bar{T}_v^\mu {\cal A}_\mu^{} B_v^{} 
                   + \bar{B}_v^{} {\cal A}_\mu^{} T_v^\mu \right)    
+ 2{\cal H}\; \bar{T}_v^\mu\, S_v^{}\cdot{\cal A}\, T_{v\mu}^{}   \;,       
\end{eqnarray}      
where  $\Delta m$  denotes the mass difference between the decuplet 
and octet baryons in the chiral-symmetry limit, $S_v^\mu$ is the 
velocity-dependent spin operator of Ref.~\cite{manjen},  and    
$\, {\cal A}_\mu^{}=\ratio{{\rm i}}{2}
\bigl( \xi\, \partial_\mu^{}\xid-\xid\,\partial_\mu^{}\xi \bigr) .\;$

Within the standard model,  the  $\,|\Delta S|=1$,  
$\,|\Delta\bfm{I}|=3/2\,$  weak transitions are induced by an effective  
Hamiltonian that transforms as  $(27_{\rm L}^{},1_{\rm R}^{})$  
under chiral rotations and has a unique chiral realization in 
the baryon-octet sector at leading order in  $\chi$PT~\cite{heval}.   
Similarly, at lowest order in $\chi$PT,  there is only one 
operator with the required transformation properties involving 
two decuplet-baryon fields, and there are no operators that involve 
one decuplet-baryon and one octet-baryon fields~\cite{atv}. 
The leading-order weak chiral Lagrangian is, thus, 
\begin{eqnarray}   
{\cal L}^{\rm w}  \;=\;     
\beta_{27}^{}\, 
T_{ij,kl}^{} \left( \xi\bar{B}_v^{} \xid \right) _{\!ki}^{} 
\left( \xi B_v^{} \xid \right) _{\!lj}^{}   
\,+\,  
\delta_{27}^{}\, 
T_{ij,kl}^{}\; \xi_{kd}^{} \xi_{bi}^\dagger\; 
\xi_{le}^{} \xi_{cj}^\dagger\; 
\bigl( \bar{T}_v^\mu \bigr) _{abc}^{} 
\bigl( T_{v\mu}^{} \bigr) _{ade}^{} 
\;+\;  {\rm h.c.}   
\label{loweak}
\end{eqnarray}      
The non-zero elements of $T_{ij,kl}^{}$ that project out the  
$\,|\Delta S|=1$,  $\,|\Delta\bfm{I}|=3/2\,$  Lagrangian are 
$\,T_{12,13}^{}=T_{21,13}^{}=T_{12,31}^{}=T_{21,31}^{}=1/2\,$ 
and  $\,T_{22,23}^{}=T_{22,32}^{}=-1/2.\,$
For purely-mesonic  $\,|\Delta S|=1$,  $\,|\Delta\bfm{I}|=3/2\,$   
processes, the lowest-order weak Lagrangian can be written as    
\begin{eqnarray}   
{\cal L}^{\rm w}_\phi  \;=\;  
{G_{\rm F}^{} \over \sqrt{2}} f_{\!\pi}^4 V_{ud}^{} V^*_{us}\,g_{27}^{}\, 
T_{ij,kl}^{} \left( \partial^\mu \Sigma\,\sd \right) _{\!ki}^{}  
\left( \partial_\mu^{} \Sigma\,\sd \right) _{\!lj}^{}   
\;+\;  {\rm h.c.}  
\label{mesonwe}
\end{eqnarray}   
and the constant $g_{27}^{}$ is measured to be about 
$0.16$~\cite{phyrep}. 

It is simple to see that the only contribution from these lowest-order 
Lagrangians to  $\,\Omega^-\ra\Xi\pi\,$  decays is via kaon poles, 
of  ${\cal O}(p)$.   
The weak Lagrangian in  Eq.~(\ref{loweak})  does not contain any couplings 
for the $\Omega^-$. 
This is easy to understand in terms of isospin: since the construction 
couples two decuplet fields and has  
$\,\Delta S=1\,$  and  $\,|\Delta\bfm{I}|=3/2,\,$  
it is not possible to involve the  $\Omega^-$  which has isospin 
zero.\footnote{This is the same reason why Eq.~(\ref{loweak})  cannot 
contribute to S-wave hyperon decays that involve 
the $\Lambda$~\cite{heval,atv}.}   

Before deriving the desired higher-order Lagrangian, we remark that 
we only need one that generates the P-wave components of  
$\,\Omega^-\ra\Xi\pi.\,$  
The reason is that, experimentally, the asymmetry parameter in these decays 
is small and consistent with zero~\cite{pdb}, indicating that 
they are dominated by a P-wave.  
We will, therefore, ignore any possible D-wave in our discussion.   

To construct the next-order Lagrangian,  ${\cal O}(p)$,  we form all 
possible 27-plets with one decuplet-baryon field, one octet-baryon field 
and one pion field (that enters through ${\cal A}_\mu^{}$). 
Employing standard techniques,\footnote{See, e.g., Ref.~\cite{tdlee}.}  
we treat the combination  $\,\bar{B}_{ab}^{} {\cal A}_{cd}^{} T_{efg}^{}\,$   
as a tensor product  $\,(8\otimes 8)\otimes 10\,$   
and find five different operators that transform as  27-plets, 
two of which contain couplings that include the $\Omega^-$.   
Their irreducible representations are
\begin{eqnarray}    
I_{ab,cd}^{}  \;=\;  
\left( \epsilon_{cef}^{}\, \epsilon_{dgh}^{} 
      + \epsilon_{def}^{}\, \epsilon_{cgh}^{} \right)  
\left( \bar{\cal T}_{aeg}^{}\, T_{bfh}^{}   
      + \bar{\cal T}_{beg}^{}\, T_{afh}^{} \right)   \;,      
\end{eqnarray}      
\begin{eqnarray}    
I_{ab,cd}'  &\!\!\!=&\!\!\!    
\epsilon_{cmn}^{} \left( \bar{\cal I}_{am,do}^{}\, T_{bno}^{}    
                        + \bar{\cal I}_{bm,do}^{}\, T_{ano}^{} \right) 
+ 
\epsilon_{dmn}^{} \left( \bar{\cal I}_{am,co}^{}\, T_{bno}^{}    
                        + \bar{\cal I}_{bm,co}^{}\, T_{ano}^{} \right) 
\nonumber \\ && \!\!\! \!\!\!   
-\; 
\ratio{1}{5} \left( \delta_{ac}^{}\, {\cal O}_{bd}^{\cal I} 
                   + \delta_{bc}^{}\, {\cal O}_{ad}^{\cal I}  
                   + \delta_{ad}^{}\, {\cal O}_{bc}^{\cal I} 
                   + \delta_{bd}^{}\, {\cal O}_{ac}^{\cal I} \right)   \;,     
\end{eqnarray}      
where    
\begin{eqnarray}    
\bar{\cal T}_{abc}^{}  \;=\;  
\epsilon_{amn}^{} \left( \bar{B}_{bm}^{} {\cal A}_{cn}^{}  
                        + \bar{B}_{cm}^{} {\cal A}_{bn}^{} \right) 
+ \epsilon_{bmn}^{} \left( \bar{B}_{cm}^{} {\cal A}_{an}^{}  
                          + \bar{B}_{am}^{} {\cal A}_{cn}^{} \right) 
+ \epsilon_{cmn}^{} \left( \bar{B}_{am}^{} {\cal A}_{bn}^{}  
                          + \bar{B}_{bm}^{} {\cal A}_{an}^{} \right)   \;,   
\end{eqnarray}      
\begin{eqnarray}    
\bar{\cal I}_{ab,cd}^{}  &\!\!\!=&\!\!\!    
\bar{B}_{ac}^{} {\cal A}_{bd}^{} + \bar{B}_{ad}^{} {\cal A}_{bc}^{}  
+ \bar{B}_{bc}^{} {\cal A}_{ad}^{} + \bar{B}_{bd}^{} {\cal A}_{ac}^{}  
\nonumber \\ && \!\!\! \!\!\!   
-\; 
\ratio{1}{5} \left( 
\delta_{ac}^{} \bar{\cal D}_{bd}^{} 
+ \delta_{ad}^{} \bar{\cal D}_{bc}^{}  
+ \delta_{bc}^{} \bar{\cal D}_{ad}^{} 
+ \delta_{bd}^{} \bar{\cal D}_{ac}^{}   
\right)   
\,-\,  
\ratio{1}{6} \left( \delta_{ac}^{}\delta_{bd}^{}  
                   + \delta_{ad}^{}\delta_{bc}^{} \right) 
\bar{\cal S}   \;,     
\end{eqnarray}      
\begin{eqnarray}    
{\cal O}_{ab}^{\cal I}  \;=\;  
\epsilon_{bmn}^{}\, \bar{\cal I}_{am,op}^{}\, T_{nop}^{}   
\;, \hspace{2em}     
\bar{\cal S}  \;=\;  \tr \left( \bar{B}{\cal A} \right)     
\;, \hspace{2em}     
\bar{\cal D}  \;=\;  \left\{ \bar{B}, {\cal A} \right\}    
                - \ratio{2}{3} \tr \left( \bar{B}{\cal A} \right)   \;.   
\end{eqnarray}      
The tensor  $I_{ab,cd}^{}$  satisfies  the symmetry relation  
$\,I_{ab,cd}^{}=I_{ba,cd}^{}=I_{ab,dc}^{}=I_{ba,dc}^{}\,$  and    
the tracelessness condition  $\,I_{ab,cb}^{}=0,\,$  
as does  $I_{ab,cd}'$.  
With these building blocks, the Lagrangian that transforms as  
$(27_{\rm L}^{},1_{\rm R}^{})$  and generates   
$\,\Delta S=1,$  $\,|\Delta\bfm{I}|=3/2\,$  transitions including 
$\Omega^-$ fields can be written as   
\begin{eqnarray}   \label{l1weak} 
{\cal L}_1^{\rm w}  \;=\;   
T_{ij,kl}^{}\; \xi_{ka}^{} \xi_{lb}^{} 
\left( {\cal C}_{27}^{}\, I_{ab,cd}^{} + {\cal C}_{27}'\, I_{ab,cd}' \right)   
\xi_{ci}^\dagger \xi_{dj}^\dagger \;.     
\end{eqnarray}      
This Lagrangian contains the terms
\begin{eqnarray}    
{\cal L}_{\Omega^- B\phi}^{\rm w}  &\!\!\!=&\!\!\!       
{{\cal C}_{27}^{}\over f}\, 6
\left( -\sqrt{2}\, \bar{\Sigma}_v^-\, \partial^\mu K^0 
      + 2\, \bar{\Sigma}_v^0\, \partial^\mu K^+ 
      - 2\, \bar{\Xi}_v^-\, \partial^\mu \pi^0 
      + \sqrt{2}\, \bar{\Xi}_v^0\, \partial^\mu \pi^+ \right) \Omega_{v\mu}^-
\nonumber \\ && \!\!\! \!\!  
+\;  
{{\cal C}_{27}'\over f}\, 2
\left( \sqrt{2}\, \bar{\Sigma}_v^-\, \partial^\mu K^0 
      - 2\, \bar{\Sigma}_v^0\, \partial^\mu K^+ 
      - 2\, \bar{\Xi}_v^-\, \partial^\mu \pi^0 
      + \sqrt{2}\,\bar{\Xi}_v^0\,\partial^\mu\pi^+ \right) \Omega_{v\mu}^-   \;. 
\end{eqnarray}      

From this expression, one can see that the decay modes  
$\,\Omega^-\ra\Xi\pi\,$  measure the combination  
$\,3{\cal C}_{27}+{\cal C}^\prime_{27}.\,$   
Since the decays  $\,\Omega^-\ra\Sigma K\,$  are
kinematically forbidden, and since three body decays of the $\Omega^-$
are poorly measured, it is not possible at present to extract 
these two constants separately.

\section{$|\Delta\bfm{I}|=3/2$ Amplitudes for $\Omega^-\ra\Xi\pi$  Decays}
  
In the heavy-baryon formalism, we can write the amplitudes as
\begin{eqnarray}   \label{amplitude}     
\ri {\cal M}_{\Omega^-\rightarrow\Xi\pi}^{}  \;=\;  
G_{\rm F}^{} m_{\pi}^2\; \bar{u}_\Xi^{}\, 
{\cal A}_{\Omega^-\Xi\pi}^{\rm (P)}\, k_\mu^{}\, u_\Omega^\mu   
\;\equiv\;  
G_{\rm F}^{} m_{\pi}^2\; \bar{u}_\Xi^{}\, 
{\alpha_{\Omega^-\Xi}^{\rm (P)}\over \sqrt{2}\, f}\,  
k_\mu^{}\, u_\Omega^\mu      \;,
\end{eqnarray}    
where  the  $u$'s  are baryon spinors,  $k$  is the outgoing four-momentum 
of the pion, and only the dominant P-wave piece is included. 
The  $\,|\Delta\bfm{I}|=3/2\,$  amplitudes satisfy the isospin relation  
$\,{\cal M}_{\Omega^-\rightarrow\Xi^-\pi^0}^{}  
   + \sqrt{2} {\cal M}_{\Omega^-\rightarrow\Xi^0\pi^-}^{} = 0 .\,$

Summing over the spin of the  $\Xi$  and  averaging over the spin of 
the  $\Omega^-$,  one derives from  Eq.~(\ref{amplitude})  
the decay width  
\begin{eqnarray}   \label{width''}     
\Gamma(\Omega^-\ra\Xi\pi)  \;=\;  
{|\bfm{k}| m_{\Xi}^{}\over 6\pi m_{\Omega}^{}} 
\left[ \left( m_{\Omega}^{}-m_\Xi^{} \right) ^2-m_\pi^2 \right]    
\Bigl| {\cal A}_{\Omega^-\Xi\pi}^{\rm (P)} \Bigr|^2 \, 
G_{\rm F}^2 m_{\pi}^4   \;.  
\end{eqnarray}    
Using the measured decay rates~\cite{pdb} and  isospin-multiplet average 
masses, we obtain the amplitudes  
\begin{eqnarray}         
{\cal A}^{({\rm P})}_{\Omega^-\Xi^-\pi^0}  \;=\;   
(3.31\pm 0.08) \;{\rm GeV}^{-1}  
\;, \hspace{3em}  
{\cal A}^{({\rm P})}_{\Omega^-\Xi^0\pi^-}  \;=\;   
(5.48\pm 0.09) \;{\rm GeV}^{-1}   \;,     
\end{eqnarray}   
up to an overall sign, where 
the relative sign between the amplitudes is chosen so that 
the $\,|\Delta\bfm{I}|=1/2\,$ rule is approximately satisfied. 
Upon defining the  $\,|\Delta\bfm{I}|=1/2,\, 3/2\,$  amplitudes 
\begin{eqnarray}   \label{a1,a3}      
\alpha^{(\Omega)}_1  \;\equiv\; 
\ratio{1}{\sqrt{3}} 
\left( \alpha^{({\rm P})}_{\Omega^-\Xi^-} 
      + \sqrt{2}\, \alpha^{({\rm P})}_{\Omega^-\Xi^0} \right)   
\;, \hspace{3em}  
\alpha^{(\Omega)}_3  \;\equiv\; 
\ratio{1}{\sqrt{3}} 
\left( \sqrt{2}\, \alpha^{({\rm P})}_{\Omega^-\Xi^-} 
      - \alpha^{({\rm P})}_{\Omega^-\Xi^0} \right)   \;,   
\end{eqnarray}   
respectively, we can extract the ratio
\begin{eqnarray}   
\alpha_{3}^{(\Omega)}/\alpha_{1}^{(\Omega)}  \;=\;  -0.072\pm 0.013   \;,
\end{eqnarray}      
which is similar to the result of Ref.~\cite{phyrep}.  
This ratio is higher than the corresponding ratios in other hyperon 
decays~\cite{atv}, which range from  $0.03$  to  $0.06$ in magnitude,  
but not significantly so.

\begin{figure}[t]         
   \hspace*{\fill} 
\begin{picture}(80,100)(-40,-35)    
\Text(-30,0)[rc]{\footnotesize $\Omega^-\,$}   
\Line(-30,1)(0,1) \Line(-30,-1)(0,-1) 
\DashLine(0,0)(0,40){3} \Text(0,45)[b]{\footnotesize $\pi$}   
\Line(0,0)(30,0) \Text(30,0)[lc]{\footnotesize $\,\Xi$}     
\Text(0,-25)[c]{(a)}   
\SetWidth{1} \BBoxc(0,0)(5,5)         
\end{picture}   
   \hspace*{\fill} 
\begin{picture}(80,100)(-40,-35)    
\Text(-30,0)[rc]{\footnotesize $\Omega^-\,$}   
\Line(-30,1)(0,1) \Line(-30,-1)(0,-1) 
\Vertex(0,0){3} \Line(0,0)(30,0) \Text(30,0)[lc]{\footnotesize $\,\Xi$}     
\DashLine(0,0)(0,40){3} \Text(0,45)[b]{\footnotesize $\pi$}   
\Text(0,10)[lb]{\footnotesize $\,K$}   
\Text(0,-25)[c]{(b)}   
\SetWidth{1} \BBoxc(0,25)(5,5)         
\end{picture}   
   \hspace*{\fill} 
\caption{\label{tree}%
Tree-level diagrams for the $\,|\Delta\bfm{I}|=3/2\,$ amplitudes 
of the P-wave  $\,\Omega^-\rightarrow\Xi\pi\,$  decays.   
In all figures, a solid dot (hollow square) represents a strong (weak) 
vertex, and the strong vertices are generated by  
${\cal L}^{\rm s}$  in  Eq.~(\ref{L1strong}).  
Here the weak vertices come from  (a) ${\cal L}_1^{\rm w}$  in  
Eq.~(\ref{l1weak})  and  (b) ${\cal L}_\phi^{\rm w}$ 
in  Eq.~(\ref{loweak}).}
\end{figure}
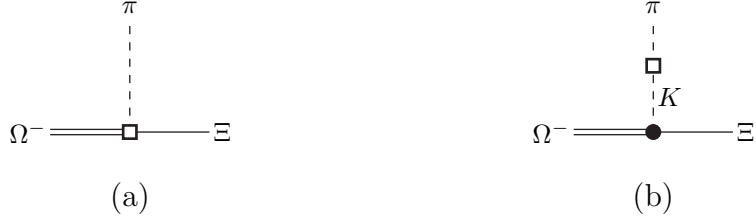

At tree level, the theoretical P-wave amplitudes arise from the diagrams 
displayed in  Figure~\ref{tree}.   
The contact diagram, Figure~\ref{tree}(a), yields  
\begin{eqnarray}   \label{contact}      
\alpha^{({\rm P})}_{\Omega^-\Xi^-}  \;=\;   
-4\sqrt{2} \left( 3{\cal C}_{27}^{} + {\cal C}_{27}' \right)  
\;, \hspace{3em}  
\alpha^{({\rm P})}_{\Omega^-\Xi^0}  \;=\;   
4 \left( 3{\cal C}_{27}^{} + {\cal C}_{27}' \right)   \;,   
\end{eqnarray}   
whereas the kaon-pole diagram, Figure~\ref{tree}(b),  gives 
\begin{eqnarray}   \label{pole}      
\alpha^{(\rm P)}_{\Omega^-\Xi^-}  \;=\;   
-2\, {\cal C} V_{ud}^{} V^*_{us} g_{27}^{}\, 
{f_{\!\pi}^2\over m_K^2-m_\pi^2}   
\;, \hspace{3em}  
\alpha^{({\rm P})}_{\Omega^-\Xi^0}  \;=\;   
\sqrt{2}\, {\cal C} V_{ud}^{} V^*_{us} g_{27}^{}\, 
{f_{\!\pi}^2\over m_K^2-m_\pi^2}   \;.       
\end{eqnarray}

The value of  the constant  $\,3{\cal C}_{27}^{}+{\cal C}_{27}'\,$   
can be extracted using the expression   
\begin{eqnarray}   \label{a3}      
\alpha^{(\Omega)}_3  \;=\;   
-4\sqrt{3} \left( 3{\cal C}_{27}^{} + {\cal C}_{27}' \right)  
\,-\, 
\sqrt{6}\, {\cal C} V_{ud}^{} V^*_{us} g_{27}^{}\, 
{f_{\!\pi}^2\over m_K^2-m_\pi^2}   \;.  
\end{eqnarray}
The kaon-pole term turns out to be small, being less than 
$10\%$  of the experimental  $\alpha^{(\Omega)}_3$, 
and so it will be neglected.   
Taking  $\,f=f_{\!\pi}^{}\approx 92.4\;\rm MeV,\,$  we then find 
\begin{eqnarray}         
3{\cal C}_{27}^{} + {\cal C}_{27}'  \;=\;  
(8.7 \pm 1.6)\times 10^{-3}\; G_{\rm F}^{} m_{\pi}^2   \;.    
\label{fitom}
\end{eqnarray}   
This value is consistent with power counting, being suppressed by 
approximately a factor of  $\Lambda_{\chi\rm SB}^{}$ with respect to the 
$\beta_{27}^{}$  found in Ref.~\cite{atv}.

\section{Octet-Hyperon Non-leptonic Decays}
  
We now address the question of the size of the contribution of  
${\cal L}_1^{\rm w}$  in  Eq.~(\ref{l1weak})  to octet-hyperon decays at 
one-loop.\footnote{%
Here, we note that the $\,|\Delta\bfm{I}|=3/2\,$ interaction,  
Eq.~(\ref{l1weak}),  does not contribute at one loop to  
$\,K\ra\pi\pi\,$  decays, and so there is no constraint from the kaon 
sector.}
There are two terms in the amplitude for the decay   
$\,B\rightarrow B^\prime\pi,\,$  corresponding to S- and P-wave 
contributions.   
In our calculation we refer exclusively to the 
$\,|\Delta\bfm{I}|=3/2\,$  component of these amplitudes. 
We follow Refs.~\cite{atv,jenkins} to write the amplitude in the form
\begin{eqnarray}        
\ri {\cal M}^{}_{B_{}\rightarrow B_{}'\pi}   \;=\;  
G_{\rm F}^{} m_{\pi}^2\, 
\bar{u}_{B_{}'}^{} \left( 
{\cal A}^{(\rm S)}_{B_{}^{}B_{}'\pi}   
+ 2 k\cdot S_v^{}\, {\cal A}^{(\rm P)}_{B_{}^{}B_{}'\pi} 
\right) u_{B_{}^{}}^{}   \;,  
\end{eqnarray}    
where  $k$  is the outgoing four-momentum of the pion. There are 
four independent amplitudes, and, as discussed in Ref.~\cite{atv}, we 
choose them to be  $\,\Sigma^+\rightarrow n\pi^+,\,$  
$\Sigma^-\rightarrow n\pi^-,\,$  $\Lambda\rightarrow p\pi^-\,$   
and  $\,\Xi^-\rightarrow\Lambda\pi^-.\,$

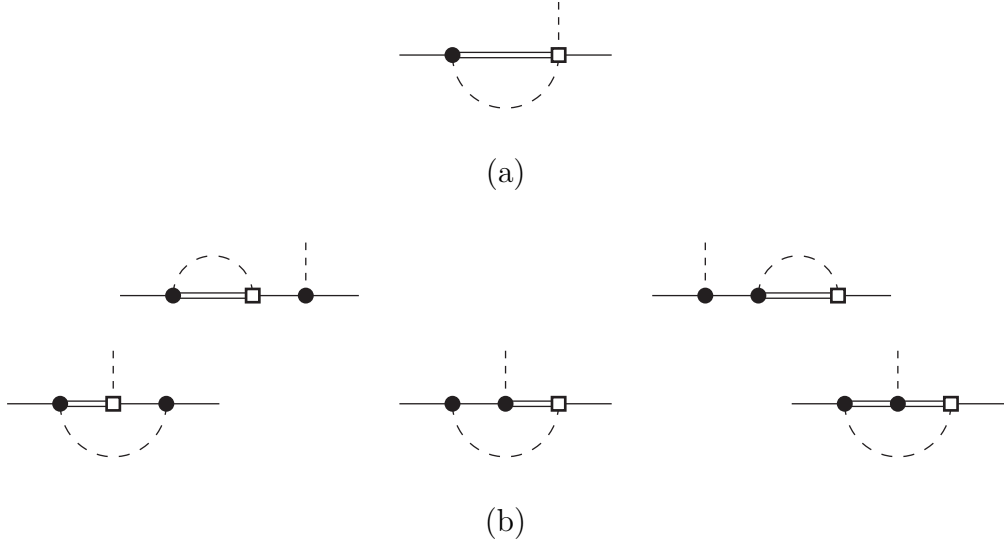
\begin{figure}[ht]         
   \hspace*{\fill} 
\begin{picture}(80,70)(-40,-50)
\Line(-40,0)(-20,0) \Vertex(-20,0){3} 
\Line(-20,1)(20,1) \Line(-20,-1)(20,-1) \Line(20,0)(40,0) 
\DashCArc(0,0)(20,180,360){4} \DashLine(20,0)(20,20){3} 
\Text(0,-45)[c]{(a)}   
\SetWidth{1} \BBoxc(20,0)(5,5) 
\end{picture}
   \hspace*{\fill} 
\\ 
   \hspace*{\fill} 
\begin{picture}(90,60)(-50,-20)   
\Line(-50,0)(-30,0) \Line(-30,1)(0,1) \Line(-30,-1)(0,-1) 
\DashCArc(-15,0)(15,0,180){4}        
\Vertex(-30,0){3} \Vertex(20,0){3} \DashLine(20,0)(20,20){3}   
\Line(0,0)(40,0) 
\SetWidth{1} \BBoxc(0,0)(5,5) 
\end{picture}
   \hspace*{\fill} 
\begin{picture}(90,60)(-40,-20)   
\Line(-40,0)(0,0) \DashLine(-20,0)(-20,20){3}  
\DashCArc(15,0)(15,0,180){4} \Vertex(-20,0){3} \Vertex(0,0){3} 
\Line(0,1)(30,1) \Line(0,-1)(30,-1) \Line(30,0)(50,0) 
\SetWidth{1} \BBoxc(30,0)(5,5) 
\end{picture}    
   \hspace*{\fill} 
\\       
   \hspace*{\fill} 
\begin{picture}(80,70)(-40,-50)
\Line(-40,0)(-20,0) \Line(-20,1)(0,1) \Line(-20,-1)(0,-1) \Line(0,0)(40,0) 
\DashCArc(0,0)(20,180,360){4}   
\Vertex(-20,0){3} \Vertex(20,0){3} 
\DashLine(0,0)(0,20){3}    
\SetWidth{1} \BBoxc(0,0)(5,5)             
\end{picture}   
   \hspace*{\fill} 
\begin{picture}(80,70)(-40,-50)
\Line(-40,0)(0,0) \Line(0,1)(20,1) \Line(0,-1)(20,-1) \Line(20,0)(40,0) 
\DashCArc(0,0)(20,180,360){4}   
\Vertex(-20,0){3} \Vertex(0,0){3} \DashLine(0,0)(0,20){3}    
\Text(0,-45)[c]{(b)}   
\SetWidth{1} \BBoxc(20,0)(5,5)             
\end{picture}
   \hspace*{\fill} 
\begin{picture}(80,70)(-40,-50)
\Line(-40,0)(-20,0) \Line(-20,1)(20,1) \Line(-20,-1)(20,-1) \Line(20,0)(40,0) 
\DashCArc(0,0)(20,180,360){4}   
\Vertex(-20,0){3} \Vertex(0,0){3} 
\DashLine(0,0)(0,20){3}    
\SetWidth{1} \BBoxc(20,0)(5,5)             
\end{picture}
   \hspace*{\fill} 
\caption{\label{loop}%
One-loop diagrams contributing to the (a) S-wave and (b) P-wave 
amplitudes of the $\,|\Delta\bfm{I}|=3/2\,$ non-leptonic decays 
of the spin-$1/2$ hyperons, 
with the weak vertices coming from  ${\cal L}_1^{\rm w}$  in  
Eq.~(\ref{l1weak}). 
A dashed line denotes an octet-meson field, and a single (double) 
solid-line denotes an octet-baryon (decuplet-baryon) field.}
\end{figure}             

Contributions of   ${\cal L}_1^{\rm w}$  to the S- and P-wave decay 
amplitudes at the one-loop level arise only from the diagrams of  
Figure~\ref{loop},  and they can be expressed in the form 
\begin{eqnarray}   \label{defswave}
{\cal A}^{\rm (S)}_{B_{}^{}B_{}'\pi}  \;=\;   
{1\over\sqrt{2}\, f_{\!\pi}^{}}\, \eta^{\rm (S)}_{B_{}^{}B_{}'}     
{m_K^3\over 24\pi f_{\!\pi}^2}   \;,   
\end{eqnarray}    
\begin{eqnarray}   \label{defpwave}   
{\cal A}^{\rm (P)}_{B_{}^{}B_{}'\pi}  \;=\;   
{1\over\sqrt{2}\, f_{\!\pi}^{}} \Biggl(\,  
\eta^{\rm (P)}_{B_{}^{}B_{}'}\, {m_K^3\over 24\pi f_{\!\pi}^2}    
\;+\;  
{\beta}^{\prime\rm (P)}_{B_{}^{}B_{}'}\,  
{m_K^2\over 16\pi^2 f_{\!\pi}^2}\, \ln{m_K^2\over\mu^2}  
\,\Biggr)   \;.
\end{eqnarray}    
Implicit in this form is the prescription of Refs.~\cite{jenkins,atv} 
in which only the non-analytic terms are kept.  
Interestingly, the only non-vanishing contribution to 
S-wave amplitudes occurs for $\Sigma$ decays and it is finite.  
Our results are   
\begin{eqnarray}   
\begin{array}{c}   \displaystyle        
\eta^{({\rm S})}_{\Lambda p}     \;=\;  
\eta^{({\rm S})}_{\Xi^-\Lambda}  \;=\;  0   \;,   
\vspace{3ex} \\   \displaystyle  
\eta^{({\rm S})}_{\Sigma^+ n}  \;=\;   
\ratio{16}{15} \bigl( 6 + \sqrt{3} \bigr) \, {\cal C}\,  
\bigl( 5 {\cal C}_{27}^{} - 3 {\cal C}_{27}' \bigr)   
\;, \hspace{3em}   
\eta^{({\rm S})}_{\Sigma^- n}  \;=\;   
\ratio{32}{45} \bigl( 6 + \sqrt{3} \bigr) \, {\cal C}\,  
\bigl( -5 {\cal C}_{27}^{} + 3 {\cal C}_{27}' \bigr)   \;, 
\end{array}   
\end{eqnarray}             
\begin{eqnarray}   
\begin{array}{c}   \displaystyle        
\eta^{({\rm P})}_{\Lambda p}  \;=\;  
\ratio{16\sqrt{2}}{45} \bigl( 1+2\sqrt{3} \bigr) \, {\cal C} D\,   
{-5{\cal C}_{27}^{} + 3{\cal C}_{27}'\over m_\Sigma^{}-m_N^{}}   \;,  
\vspace{3ex} \\   \displaystyle  
\eta^{({\rm P})}_{\Xi^-\Lambda}  \;=\;      
\ratio{8\sqrt{2}}{45}\, {\cal C} D\, 
{ 10 \bigl( 1+2\sqrt{3} \bigr) \, {\cal C}_{27}^{} 
 - \bigl( 2+7\sqrt{3} \bigr) \, {\cal C}_{27}'  \over  m_\Xi^{}-m_\Sigma^{} }  
 \;,
\vspace{3ex} \\   \displaystyle  
\eta^{({\rm P})}_{\Sigma^+ n}  \;=\;    
\ratio{16}{45} \bigl( 6+\sqrt{3} \bigr) \, {\cal C}\, (D+3 F)\,    
{5{\cal C}_{27}^{} - 3 {\cal C}_{27}'\over m_\Sigma^{}-m_N^{}}   
\;, \hspace{3em}   
\eta^{({\rm P})}_{\Sigma^- n}  \;=\;   
\ratio{32}{45} \bigl( 6+\sqrt{3} \bigr) \, {\cal C} F\, 
{-5{\cal C}_{27}^{} + 3{\cal C}_{27}'\over m_\Sigma^{}-m_N^{}}   \;,   
\end{array}   
\end{eqnarray}             
\begin{eqnarray}   
\begin{array}{c}   \displaystyle        
\beta^{\prime({\rm P})}_{\Lambda p}  \;=\;   
\ratio{-4}{27\sqrt{6}}\, {\cal C}\, 
\bigl[ (54 D + 162 F + 5 {\cal H})\, {\cal C}_{27}^{}    
      - (90 D + 54 F + {\cal H})\, {\cal C}_{27}' \bigr]   \;,  
\vspace{3ex} \\   \displaystyle  
\beta^{\prime({\rm P})}_{\Xi^-\Lambda}  \;=\;  
\ratio{-4}{135\sqrt{6}}\, {\cal C}\, 
\bigl[ (270 D - 810 F - 25 {\cal H})\, {\cal C}_{27}^{}
      + (54 D - 234 F - 15 {\cal H})\, {\cal C}_{27}' \bigr]   \;,   
\vspace{3ex} \\   \displaystyle  
\beta^{\prime({\rm P})}_{\Sigma^+ n}  \;=\;  
\ratio{8}{135}\, {\cal C}\, 
\bigl[ (10 D - 125 {\cal H})\, {\cal C}_{27}^{}  
      + (-86 D + 84 F + 55 {\cal H})\, {\cal C}_{27}' \bigr]   \;,  
\vspace{3ex} \\   \displaystyle  
\beta^{\prime({\rm P})}_{\Sigma^- n}  \;=\;  
\ratio{4}{27}\, {\cal C}\, 
\bigl[ (-62 D + 54 F + 35 {\cal H}) \, {\cal C}_{27}^{} 
      + (18 D - 18 F - 27 {\cal H}) \, {\cal C}_{27}' \bigr]   \;.  
\end{array} 
\end{eqnarray}             

\section{Results and Conclusion}

The contributions from  Eq.~(\ref{l1weak})  to octet-baryon non-leptonic
decay can be summarized numerically in terms of ${\cal C}_{27}^{}$  
and  ${\cal C}_{27}'$ as follows:   
\begin{eqnarray}   
\begin{array}{c}   \displaystyle        
S_3^{(\Lambda)}  \;=\;  S_3^{(\Xi)}  \;=\;  0   \;, \hspace{3em} 
S_3^{(\Sigma)}   \;=\;  -100.6\, {\cal C}_{27}^{} + 60.38\, {\cal C}_{27}'   
\;,
\vspace{3ex} \\   \displaystyle  
P_3^{(\Lambda)}  \;=\;   8.089\, {\cal C}_{27}^{} - 4.127\, {\cal C}_{27}'    
\;, \hspace{3em}  
P_3^{(\Xi)}      \;=\;  -33.62\, {\cal C}_{27}^{} + 11.34\, {\cal C}_{27}'   
\;,
\vspace{3ex} \\   \displaystyle  
P_3^{(\Sigma)}   \;=\;   18.64\, {\cal C}_{27}^{} - 10.42\, {\cal C}_{27}'   
\;.
\end{array}        
\end{eqnarray}       
Here, we have employed the parameter values
$\,D=0.61,$  $\,F=0.40,$  $\,{\cal C}=1.6,$  and  $\,{\cal H}=-1.9,\,$  
obtained in Ref.~\cite{manjenco}.   
The measured rates for  $\,\Omega^-\ra\Xi\pi\,$  only
determine the combination  $\,3 {\cal C}_{27}^{}+{\cal C}_{27}',\,$ 
as indicated in  Eq.~(\ref{fitom}). 
As an illustration of the effect of these terms on the octet-hyperon 
non-leptonic decay,  we present numerical results in Table~\ref{result1}, 
where  we look at four simple scenarios to 
satisfy  Eq.~(\ref{fitom})  in terms of only one parameter. 
\begin{table}[b]      
\caption{\label{result1}%
New $\;|\Delta\bfm{I}|=3/2\;$  contributions to 
S- and P-wave hyperon decay  amplitudes compared with experiment and
with the best theoretical fit of Ref.~\cite{atv}.  
Here  ${\cal C}_{27}^{}$  and  ${\cal C}_{27}'$  are given in units 
of  $\;10^{-3}\;G_{\rm F}^{} m_{\pi}^2,\;$  and their values are chosen 
to fit the  $\;\Omega^-\ra\Xi\pi\;$  decays.}
\centering   \small 
\vskip 1\baselineskip  
\begin{tabular}{crrrrrrrrrrrr}    
\hline \hline      
\vspace{-2ex} \\  
&& Theory\hspace{-1ex} && 
\multicolumn{9}{c}{Theory, new contributions with  
$\;3{\cal C}_{27}^{}+{\cal C}_{27}'=8.7\;$} 
\vspace{1ex} \\ \cline{5-13}    
\raisebox{2ex}[2ex]{Amplitude}  &  
\raisebox{2ex}[2ex]{Experiment\hspace{1ex}}  &  
\raisebox{0.5ex}{Ref.~\cite{atv}\hspace{-1ex}} &&&  
\multicolumn{2}{c}{${\cal C}_{27}'=0$}\hspace{3ex} &
\multicolumn{2}{c}{${\cal C}_{27}^{}=0$}\hspace{3ex} & 
${\cal C}_{27}'={\cal C}_{27}^{}$\hspace{-2.5ex} &&  
\hspace{2ex}${\cal C}_{27}'=-{\cal C}_{27}^{}$\hspace{-4ex} & 
\vspace{0.3ex} \\ \hline      
\vspace{-2.5ex} \\  
$S_3^{(\Sigma)}$   &   $-0.107\pm 0.038$  & $-$0.120  &&&  $-$0.29   &&  
0.52               &&  $-$0.09   && $-$0.70 &         
\\  
$P_3^{(\Lambda)}$  &   $-0.021\pm 0.025$  & $-$0.023  &&& 0.02       &&  
$-$0.04            &&  0.01      && 0.05  &      
\\  
$P_3^{(\Xi)}$      &   $ 0.022\pm 0.023$  & 0.027     &&& $-$0.10    &&   
0.10               &&  $-$0.05   && $-$0.20 &           
\\  
\vspace{.3ex}   
$P_3^{(\Sigma)}$   &   $-0.110\pm 0.045$  & $-$0.066  &&& 0.05       &&   
$-$0.09            &&  0.02      && 0.13  & \hspace*{1em}      
\\  
\hline \hline 
\end{tabular}   
\end{table}
For comparison, we show in the same Table the experimental value of 
the amplitudes as well as the best theoretical fit at  
${\cal O}(m_s\log m_s)$  obtained in Ref.~\cite{atv}. 
The new terms calculated here (with  $\,\mu=1\;\rm GeV$), induced by  
${\cal L}_1^{\rm w}$  in  Eq.~(\ref{l1weak}),  are of higher order in  
$m_s^{}$  and  are therefore expected to be smaller than the best 
theoretical fit.  
A quick glance at  Table~\ref{result1}  shows that in some cases the new 
contributions are much larger.  
Another way to gauge the size of the new contributions is to compare them  
with the experimental error in the octet-hyperon decay amplitudes.  
Since the theory provides a good fit at  ${\cal O}(m_s\log m_s)$~\cite{atv},  
we would like the new contributions (which are of higher order in  $m_s^{}$)  
to be at most at the level of the experimental error.  
From  Table~\ref{result1},  we see that in some cases the new 
contributions are significantly larger than these errors. 
In a few cases they are significantly larger than the experimental 
amplitudes.  
All this indicates to us that the measured  $\,\Omega^-\ra\Xi\pi\,$  
decay rates imply a  $\,|\Delta\bfm{I}|=3/2\,$  amplitude that may be 
too large and in contradiction with the  $\,|\Delta\bfm{I}|=3/2\,$  
amplitudes measured in octet-hyperon non-leptonic decays.

Nevertheless, it is premature to conclude that the measured values for
the  $\,\Omega^-\ra\Xi\pi\,$  decay rates must be incorrect because,
strictly speaking, none of the contributions to octet-baryon decay
amplitudes is proportional to the same combination of parameters
measured in  $\,\Omega^-\ra\Xi\pi\,$  decays,  
$\,3 {\cal C}_{27}^{}+{\cal C}_{27}'.\,$  
It is possible to construct linear combinations of the four amplitudes  
$S_3^{(\Sigma)}$,  $P_3^{(\Sigma)}$,  $P_3^{(\Lambda)}$  and  $P_3^{(\Xi)}$  
that are proportional to  $\,3 {\cal C}_{27}^{}+{\cal C}_{27}'.\,$
We find that the most sensitive one is   
\begin{equation}
\left( S_3^{(\Sigma)}-4.2 P_3^{(\Xi)} \right) _{\rm Exp}^{}   \;=\;  
-0.2 \pm 0.1   \;,  
\end{equation}
where we have simply combined the errors in quadrature.  
The contribution from Eq.~(\ref{l1weak})  to this combination is 
\begin{equation}
\left( S_3^{(\Sigma)}-4.2 P_3^{(\Xi)} \right) _{\rm Theory,new}^{}  
\;\approx\; 
13\, \bigl( 3 {\cal C}_{27}^{}+{\cal C}_{27}' \bigr)  
\;\approx\;  0.1   \;,
\end{equation}  
which falls within the error in the measurement.

Our conclusion is that the current measurement of the rates for 
$\,\Omega^-\ra\Xi\pi\,$  implies a $\,|\Delta\bfm{I}|=3/2\,$ amplitude that 
appears large enough to be in conflict with measurements of  
$\,|\Delta\bfm{I}|=3/2\,$  amplitudes in octet-baryon non-leptonic decays. 
However, within current errors and without any additional assumptions 
about the relative size of  ${\cal C}_{27}^{}$  and  ${\cal C}_{27}'$,  
the two sets of measurements are not in conflict.

\vspace{1in}

\noindent {\bf Acknowledgments} This work  was supported in
part by DOE under contract number DE-FG02-92ER40730. 
We thank the theory group at Fermilab for their hospitality while 
part of this work was done. 
We also thank Xiao-Gang~He, K.~B.~Luk and Sandip Pakvasa for conversations, 
and W.~Bardeen for interesting discussions on 
the  $\,|\Delta\bfm{I}|=1/2\,$  rule.

\end{document}